\documentclass{article}
\usepackage{spconf,amsmath,amssymb,graphicx,hyperref}
\usepackage[utf8]{inputenc}
\usepackage[T1]{fontenc}
\usepackage{cite}
\usepackage{booktabs}
\usepackage{multirow}
\usepackage{array}
\usepackage{comment}
\usepackage{url}
\usepackage{xcolor}
\usepackage{tikz}
\usetikzlibrary{arrows.meta,positioning,fit,shapes.geometric,backgrounds,calc}

\hypersetup{
    colorlinks=true,
    linkcolor=blue!60!black,
    citecolor=blue!60!black,
    urlcolor=blue!60!black,
}

\definecolor{audiocol}{RGB}{0,121,107}     
\definecolor{imagecol}{RGB}{198,124,0}      
\definecolor{textcol}{RGB}{79,91,160}       
\definecolor{s1col}{RGB}{55,90,140}         
\definecolor{s2col}{RGB}{45,120,110}        
\definecolor{s3col}{RGB}{110,70,140}        
\definecolor{ink}{RGB}{38,42,48}

\title{Audio--Image Alignment as a Continued-Pretraining Stage in Automatic Speech Recognition}
\name{Sujith Pulikodan$^{1}$, \ Nihar Desai$^{1}$, \ Prasanta Kumar Ghosh$^{2}$}
\address{$^{1}$AI \& Robotics Technology Park (ARTPARK), I-Hub @ IISc, Bangalore, India \\
	$^{2}$Department of Electrical Engineering, Indian Institute of Science, Bangalore, India \\
	\texttt{\{sujith, nihar\}@artpark.in}, \ \texttt{prasantg@iisc.ac.in}}
\begin{document}
%
\maketitle
\begin{abstract}
Thousands of languages are spoken worldwide, yet many remain under-resourced
for Automatic Speech Recognition (ASR) due to the limited availability of
high-quality transcribed speech data. Collecting accurate transcriptions is
often costly and labor-intensive, particularly for low-resource languages. In this work, we introduce a representation-alignment stage between
large-scale pretraining and supervised ASR fine-tuning, in which image
representations extracted from pretrained vision encoders are aligned with
audio representations to adapt a pretrained audio encoder using only paired
audio--image data, with no transcriptions. On FLEURS, alignment reduces overall WER across 11 Indian languages from 66.42\%
to 49.77\%, a 25.07\% relative reduction that is statistically significant in
every language. Two controls establish that the gain comes from the image
signal itself: a compute-matched continued-pretraining baseline yields no
improvement (66.42\% vs.\ 65.99\%), and shuffling the audio--image pairs
removes the benefit entirely (72.60\%). The aligned encoder also improves over
the baseline at every fine-tuning budget from 10 to 100 hours on Vaani and
LibriSpeech, with the largest gains in the lowest-resource settings. These
findings highlight audio--image representation alignment as an effective
transcription-free adaptation strategy for ASR.
\end{abstract}
\begin{keywords}
ASR, audio--image alignment, multimodal representation learning
\end{keywords}
%
\section{Introduction}
Pretraining and fine-tuning have become the dominant paradigm for developing AI models, including Automatic Speech Recognition (ASR) systems. Audio encoders are typically pretrained on large-scale unlabeled speech and fine-tuned using labeled audio--text pairs \cite{wav2vec2,hubert}. While this approach effectively leverages unlabeled speech, many languages remain limited by the availability of transcribed data, as high-quality transcription is expensive, time-consuming, and requires language expertise. Image-prompted speech collection, used at scale in corpora such as
WAXAL~\cite{waxal} and Vaani~\cite{Vaani}, elicits natural, spontaneous speech
without scripted prompts. It also yields, as a by-product, large volumes of
semantically aligned audio--image pairs --- a source of multimodal supervision
that comes with no additional annotation cost.

Motivated by this observation, we investigate whether such multimodal data can be used to further adapt pretrained audio encoders without transcription labels. Specifically, we introduce an intermediate representation-alignment stage between audio pretraining and supervised fine-tuning, where audio representations are aligned with semantic representations extracted from images. By leveraging the complementary semantic information in images, this adaptation provides a transcription-free learning signal to improve speech representations and downstream ASR performance.

\section{Related Work}

Audio-Visual Speech Recognition (AVSR) has been extensively studied as a multimodal approach to speech recognition \cite{ivanko2023review}. These systems combine audio with visual information, such as lip movements or video frames, to improve recognition, particularly in noisy environments. Related work has also incorporated scene-level visual context alongside audio for speech recognition \cite{gupta2017context}. In most cases, visual information is used during inference to improve ASR robustness.

Several self-supervised learning (SSL) approaches have been proposed for
learning speech representations from large-scale unlabeled audio, including
wav2vec 2.0~\cite{wav2vec2}, HuBERT~\cite{hubert}, and Best-RQ~\cite{bestrq}.
Similarly, vision--language models such as CLIP~\cite{clip} and
SigLIP~\cite{zhai2023sigmoid} demonstrate the effectiveness of cross-modal
representation alignment. Harwath et al.\ learned
semantic correspondences between spoken descriptions and images from
untranscribed speech, with the images providing the supervision
\cite{harwath2015deep, harwath2016unsupervised}, yielding shared audio--image
representations for visual grounding and cross-modal retrieval. Subsequent
approaches, including SpeechCLIP\cite{speechclip} and AudioCLIP\cite{audioclip}, further explored shared
representations across speech, audio, and visual modalities. In contrast, our work uses audio--image alignment
as an intermediate adaptation stage for a pretrained ASR encoder, rather than
as the final multimodal representation or inference objective.

In this work, we investigate audio--image representation alignment as an intermediate adaptation stage for pretrained ASR encoders. Unlike prior multimodal approaches, the aligned images provide a transcription-free learning signal before conventional supervised ASR fine-tuning.

\section{Approach}
\label{sec:method}

We insert an audio--image alignment stage between self-supervised pretraining
and supervised fine-tuning. A frozen vision encoder embeds each prompt image;
the audio encoder, initialized from the self-supervised checkpoint, is trained
contrastively to place each utterance near the image that elicited it and away
from the other images in the batch. A projection head maps audio into the image
embedding space and is discarded afterwards, so the encoder handed to
fine-tuning is architecturally identical to the baseline, and the audio comes
from the pretraining corpus, so the stage adds no unseen speech --- only
images. Pretraining shapes the encoder around short-span acoustic structure,
whereas alignment supplies an utterance-level target carrying no acoustics,
only a space organised by meaning; we hypothesise that this regularises
representations fine-tuning would otherwise learn from transcriptions alone, so
the benefit should be largest where transcribed data is scarce.

\subsection{Datasets}

We use the Vaani dataset for both self-supervised pretraining and multimodal audio--image alignment. The Vaani corpus comprises approximately 31,255 hours of speech spanning 105 languages, of which 2,043 hours are transcribed. Data are collected using a picture-prompt paradigm, where speakers are shown an image and asked to describe it or speak about whatever the image evokes, naturally yielding semantically aligned audio--image pairs suitable for multimodal representation learning. For self-supervised pretraining, we use the entire Vaani speech corpus (train 28,418.12 h), excluding validation and evaluation splits. For audio--image alignment, we use 11,848,593 audio--image pairs associated with approximately 287K unique district-specific images, totaling 16,580.36 hours of speech. This subset is drawn exclusively from the district-specific image portion of the Vaani dataset. These pairs are drawn entirely from the pretraining corpus and therefore do not introduce additional or previously unseen audio data. For supervised fine-tuning, we conduct three independent experiments using the transcribed portions of Vaani, FLEURS \cite{conneau2023fleurs} and LibriSpeech\cite{librispeech}, respectively, with each experiment using one dataset independently to evaluate fine-tuning across different sources of transcribed speech.

\subsection{Models}

For the audio encoder, we use a 17-layer FastConformer
architecture~\cite{fastconformer} with 432.87M parameters, first pretrained
on the Vaani corpus using self-supervised learning. The pretrained encoder initializes both the audio--image alignment stage and the baseline ASR system. For alignment, we add an MLP head (5.26M parameters) that projects audio
representations into a shared embedding space with the image representations,
and jointly optimize the encoder and alignment head. After alignment, the head
is discarded before supervised fine-tuning. We experiment with SigLIP2 Base\cite{siglip2}, SigLIP2 Large \cite{siglip2}, and Qwen3-VL \cite{qwenvl} as image encoders. The aligned audio encoder is then coupled with a hybrid CTC-TDT decoder\cite{ctc,tdt} and fine-tuned on multilingual transcribed speech. As a baseline, the original self-supervised encoder, without audio--image alignment, is coupled with the same decoder and fine-tuned under the identical setup. We use a BPE tokenizer \cite{sentencepiece} with a vocabulary size of 2,000, trained separately on the training split of each fine-tuning dataset; thus, separate tokenizers are used for the Vaani, LibriSpeech and FLEURS fine-tuning setups.

\begin{figure*}[t]
    \centering
    \includegraphics[width=0.95\linewidth]{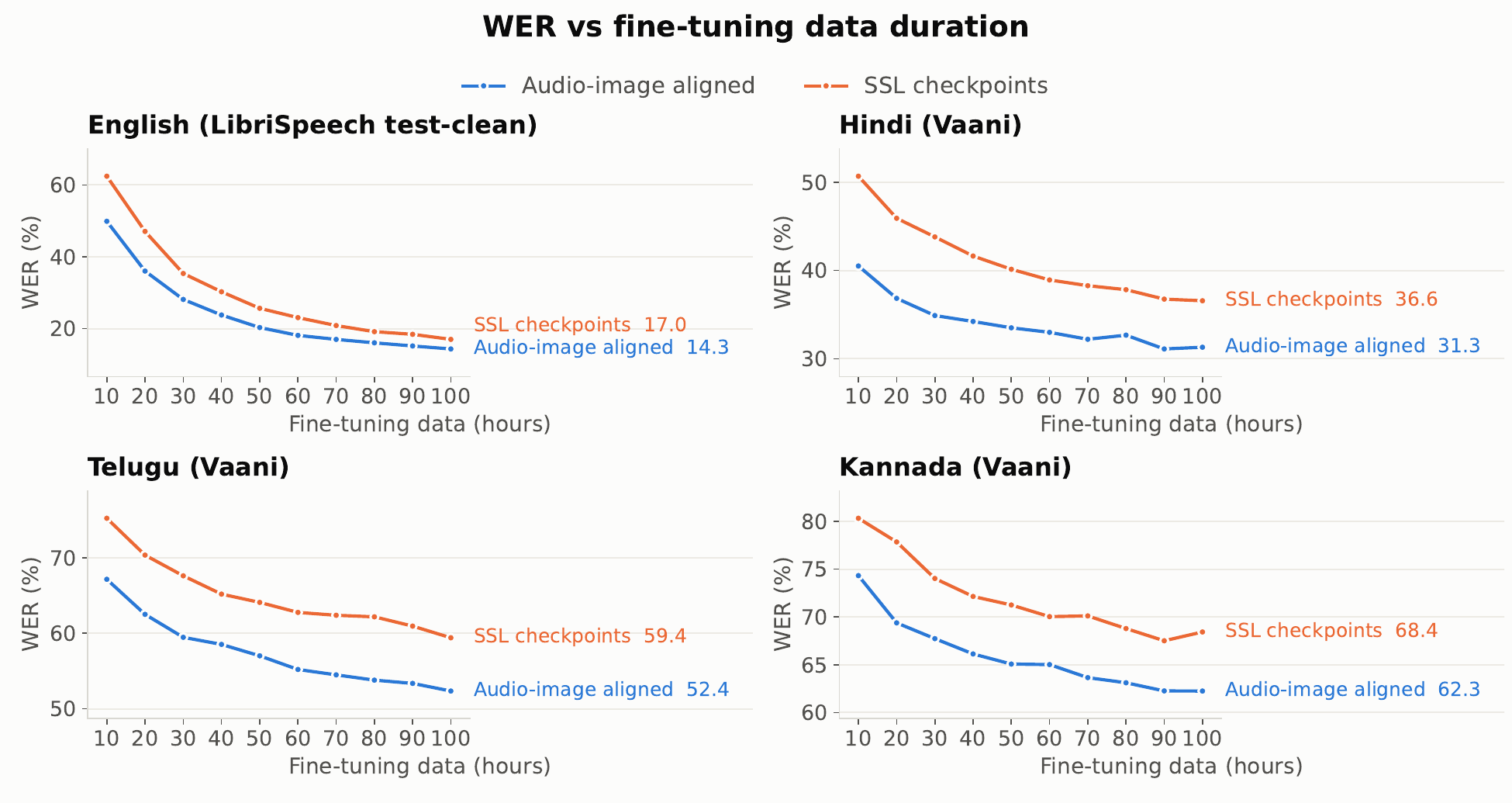}
    \caption{WER vs Fine-tuning duration }
    \label{fig:example}
\end{figure*}

\begin{table}[t]
\centering
\small
\setlength{\tabcolsep}{6pt}
\renewcommand{\arraystretch}{1.1}
\caption{Configuration of the three alignment variants.}
\label{tab:variants}
\begin{tabular}{lccc}
\toprule
 & \textbf{\shortstack{SigLIP2\\Base}}
 & \textbf{\shortstack{SigLIP2\\Large}}
 & \textbf{\shortstack{Qwen3-VL\\Embed.-2B}} \\
\midrule
Vision backbone      & \shortstack{SigLIP2\\base/256} & \shortstack{SigLIP2\\large/384} & \shortstack{Qwen3-VL\\Emb.-2B} \\
Backbone params      & $\sim$86M  & $\sim$303M & $\sim$500M \\
Embed dim $D$        & 768        & 1024       & 2048 \\
Img.\ tokens $K_v$   & \shortstack{1\\(pooled)} & \shortstack{16\\(top-$K$)} & \shortstack{16\\(merged)} \\
Audio queries $K_a$  & 1          & 16         & 16 \\
Pair score           & cosine     & MaxSim     & MaxSim \\
\bottomrule
\end{tabular}
\end{table}

\begin{table*}[!t]
\centering
\small
\setlength{\tabcolsep}{5pt}
\small
\caption{FLEURS WER (\%) across 11 Indian languages. Base ckpt is the SSL checkpoint at epoch 67; Compute-m. is the compute-matched SSL checkpoint at epoch 72. $\Delta$, Rel., 95\% CI, and $p$ compare SigLIP2 Large with the compute-matched baseline.}
\begin{tabular}{lrrrrrrrrcc}
\toprule
 &  & \multicolumn{2}{c}{SSL baseline} & \multicolumn{3}{c}{Audio--Image alignment} & \multicolumn{4}{c}{SigLIP2 Large vs.\ SSL$_\text{cm}$} \\
\cmidrule(lr){3-4}\cmidrule(lr){5-7}\cmidrule(lr){8-11}
Language & $N$ (hours) & Base ckpt & Compute-m. & Qwen3-VL & SigLIP2 L & SigLIP2 B & $\Delta$ & Rel.\,(\%) & 95\% CI on $\Delta$ & $p$ \\
\midrule
Marathi & 1,015 (3.9) & 70.22 & 71.91 & 53.49 & \textbf{52.13} & 52.20 & 19.78 & 27.51 & [18.72,\,20.86] & $<10^{-3}$ \\
Gujarati & 1,000 (2.9) & 56.97 & 56.22 & 46.41 & 46.93 & \textbf{46.11} & 9.29 & 16.53 & [8.36,\,10.21] & $<10^{-3}$ \\
Assamese & 984 (3.5) & 62.95 & 63.04 & 52.71 & \textbf{51.88} & 53.32 & 11.16 & 17.70 & [10.38,\,11.91] & $<10^{-3}$ \\
Malayalam & 958 (3.9) & 80.04 & 79.12 & 50.31 & \textbf{49.81} & 49.86 & 29.30 & 37.04 & [27.75,\,30.86] & $<10^{-3}$ \\
Bengali & 920 (3.4) & 45.19 & 45.73 & 38.51 & \textbf{37.64} & 37.69 & 8.08 & 17.68 & [7.40,\,8.78] & $<10^{-3}$ \\
Oriya & 883 (3.0) & 81.73 & 82.16 & 62.29 & \textbf{61.18} & 61.37 & 20.98 & 25.53 & [19.68,\,22.26] & $<10^{-3}$ \\
Kannada & 838 (3.2) & 79.42 & 83.61 & 47.44 & \textbf{47.33} & 47.78 & 36.28 & 43.39 & [34.60,\,38.00] & $<10^{-3}$ \\
Tamil & 591 (2.1) & 77.76 & 78.07 & 65.76 & 66.41 & \textbf{65.40} & 11.66 & 14.94 & [10.38,\,12.95] & $<10^{-3}$ \\
Punjabi & 574 (1.8) & 62.04 & 61.15 & \textbf{45.62} & 46.41 & 48.19 & 14.74 & 24.10 & [13.23,\,16.27] & $<10^{-3}$ \\
Telugu & 472 (1.4) & 64.81 & 65.23 & \textbf{54.45} & 54.69 & 54.76 & 10.54 & 16.16 & [9.26,\,11.85] & $<10^{-3}$ \\
Hindi & 418 (1.3) & 49.22 & 49.18 & 37.08 & \textbf{36.64} & 37.09 & 12.54 & 25.50 & [11.03,\,14.13] & $<10^{-3}$ \\
\midrule
\textbf{Overall} & 8,653 (30.5) & 65.99 & 66.42 & 50.13 & \textbf{49.77} & 50.03 & 16.65 & 25.07 & [16.25,\,17.06] & $<10^{-3}$ \\
\bottomrule
\end{tabular}

\label{tab:fleurs}
\end{table*}

\subsection{Image alignment}
We experimented with three audio–image alignment configurations using SigLIP2 Base, SigLIP2 Large, and Qwen3-VL as the image encoders. In all three, the audio encoder is a FastConformer self-supervised model that is fully fine-tuned during alignment, while the image encoder is kept frozen and its embeddings are precomputed and cached. The models are trained with the  sigmoid contrastive loss\cite{zhai2023sigmoid} with a learnable temperature and bias, using in-batch negatives gathered across GPUs.

In the first configuration (SigLIP2 Base, patch16-256), audio encoder outputs are pooled with a single-query attention pooling layer and projected to the 768-d SigLIP image space with an MLP. The image encoder produces one pooled embedding per image. Pair similarity is the cosine similarity between the projected audio and image embeddings. In the second configuration (SigLIP2 Large, patch16-384), the audio encoder outputs are processed with multi-query attention pooling followed by an MLP to produce 16 audio embeddings of dimension 1024. On the image side, the top 16 patch tokens (out of 576) are selected by L2 norm and L2-normalized. The pair score is computed with an asymmetric MaxSim\cite{colbert}: for each audio query we take the maximum cosine similarity over the image tokens, then average over the 16 audio queries (padded image tokens are masked). In the third configuration, the Qwen3-VL-Emb.-2B visual encoder is used; a 2×2 spatial merge yields up to 16 image tokens of dimension 2048. The audio side again uses multi-query attention pooling and an MLP to produce 16 audio embeddings, and the pair score uses the same MaxSim function as the second configuration.

All audio--image alignment models were trained using a consistent optimization setup. Training employed a batch size of 64 and the AdamW optimizer with $\beta_1 = 0.9$ and $\beta_2 = 0.95$, a weight decay of 0.01, gradient clipping at 1.0, and bfloat16 precision. Newly initialized layers were trained with a learning rate of $3 \times 10^{-4}$, while pretrained encoder parameters used a reduced learning rate scaled by a factor of 0.05. A learning rate schedule consisting of a 1{,}000-step warm-up phase followed by decay to a minimum multiplier of 0.05 was applied. All model configurations were trained for 200{,}000 optimization steps.

\subsection{Fine-Tuning }

Every fine-tuning run starts from the same architectural template ---
a hybrid CTC-TDT decoder  model with TDT durations $\{0,1,2,3,4\}$
 --- and inherits its encoder weights either from the
unaligned SSL checkpoint (\emph{baseline}) or from one of the
alignment variants. Fine-tuning is performed with NeMo \cite{nemo} on
$1\!\times\!\text{H200}$, AdamW \cite{adamw} with Noam annealing and a 2{,}000-step
warmup, batch 16 per GPU, learning rate $1\!\times\!10^{-4}$, and
SpecAugment \cite{specaugment}. All fine-tuning runs used the same early-stopping criterion: training was terminated when validation WER did not improve for four consecutive epochs. Each model therefore trains to its own validation-WER optimum rather than a fixed epoch budget. The trained models are evaluated using Word Error Rate (WER)  as the evaluation metric. We fine-tune on FLEURS covering 11 Indian languages jointly, using 93.7 hours of training data, 13.4 hours of validation data  and 30.5 hours of test data. For the data-scaling experiments, we train on nested subsets of 10 to 100 hours, in 10-hour steps, for Vaani Hindi, Kannada and Telugu and for English LibriSpeech (drawn from train-clean-100). Validation uses one tenth of the training size for Vaani (1 to 10 hours) and a fixed 5-hour subset of dev-clean for LibriSpeech. Evaluation uses the Vaani test set for Hindi (31.2 hours), Kannada (3.7 hours) and Telugu (5.0 hours), and LibriSpeech test-clean for English (5.4 hours).

\section{Results and Discussion}

We compare against two SSL-only baselines: the epoch 67 checkpoint from which
alignment starts, and an epoch 72 checkpoint obtained by continuing
self-supervised pretraining for a compute budget matched to the alignment stage. On
FLEURS, which evaluates cross-domain generalization, these reach 65.99\% and
66.42\% WER across 11 Indian languages. Alignment reduces this to 50.13\% with
Qwen3-VL, 50.03\% with SigLIP2 Base, and 49.77\% with SigLIP2 Large, a 25\%
relative reduction for the best configuration. The three image encoders differ
by less than 0.4 points despite spanning 86M--500M parameters, indicating that
the benefit comes from the alignment objective rather than from the vision
backbone. SigLIP2 Large improves over the compute-matched baseline in all 11
languages, each at $p < 10^{-3}$ under a paired utterance-level bootstrap
(Table~\ref{tab:fleurs}); per-language relative gains range from 14.9\%
(Tamil) to 43.4\% (Kannada), and are largest where the baseline is weakest.

We further evaluate the aligned encoders on image retrieval using held-out
data comprising 5,050 utterances and 3,856 images, for which chance R@1 is
0.00026. The aligned models achieve audio-to-image R@1 scores of 0.118, 0.091,
and 0.091, and image-to-audio R@1 scores of 0.172, 0.143, and 0.118 for
SigLIP2 Base, SigLIP2 Large, and Qwen3-VL, respectively --- several hundred
times chance. Retrieval quality does not, however, predict downstream ASR
gain: SigLIP2 Base retrieves best while SigLIP2 Large yields the lowest FLEURS
WER. Together with the gains being largest out-of-domain and in the
low-resource regime, this is consistent with alignment acting as a semantic
regulariser on the encoder rather than importing retrieval-specific structure.

To determine whether the ASR gains arise from genuine audio--image
correspondence rather than additional training, we aligned an identical
encoder with the audio--image pairs randomly shuffled, changing only which
image each utterance is paired with. On FLEURS this model reaches 72.60\% WER,
worse than either SSL-only baseline and far worse than the 49.77\% of the
correctly aligned encoder: a mismatched contrastive signal actively degrades
the representation, so the improvement comes from the image correspondence
rather than from additional optimization. The two SSL-only checkpoints
themselves perform almost identically (65.99\% vs.\ 66.42\%), with
per-language differences within 1.7 points except for Kannada, ruling out
checkpoint selection.

Fig.~\ref{fig:example} reports WER as a function of fine-tuning duration for
English (LibriSpeech) and Hindi, Kannada, and Telugu (Vaani); all four
languages appear in the pretraining and alignment corpora. At the largest
budget of 100 hours, the SigLIP2 Large aligned encoder reaches 14.3\% WER on
LibriSpeech test-clean against 17.0\% for the SSL-only baseline, and 31.3\%,
62.3\%, and 52.4\% on Vaani Hindi, Kannada, and Telugu against 36.6\%, 68.4\%,
and 59.4\%. It stays ahead at every duration and in every language, by
2.7--7.0 points at 100 hours, with the largest gains in the low-resource
regime narrowing as labeled data increases.
\section{Conclusion}
We demonstrate that incorporating audio--image alignment as an intermediate
continued-pretraining stage improves ASR performance without requiring any
additional transcribed speech. On FLEURS, alignment reduces WER across 11
Indian languages by 25\% relative to a compute-matched baseline, significantly
in every language, and consistent gains appear at every fine-tuning budget in
the Vaani and LibriSpeech scaling experiments. Two controls establish that the
gain comes from the audio--image correspondence itself rather than from
additional optimization: compute-matched continued pretraining yields no
improvement, and shuffling the pairs leaves the encoder worse than no
alignment at all. A key requirement of this approach is the availability of
paired audio--image data, though image-prompted collection is already used at
scale in corpora such as Vaani and WAXAL.

\vfill\pagebreak

\end{document}